\documentstyle[12pt]{article}
\def\be{\begin{equation}}
\def\ee{\end{equation}}
\def\bea{\begin{eqnarray}}
\def\eea{\end{eqnarray}}
\def\a{\alpha}
\def\b{\beta}

\def\e{\epsilon}

\author{Hans-J\"urgen Schmidt}

\title{New exact solutions for power--law inflation Friedmann models}

\date{}
\begin{document}
\maketitle

\centerline{Universit\"at Potsdam, Institut f\"ur Mathematik, Am
Neuen Palais 10} 
 \centerline{D-14469~Potsdam, Germany,  E-mail:
 hjschmi@rz.uni-potsdam.de}

\begin{abstract}
\noindent
We consider the spatially flat Friedmann model
$$
ds^2=dt^2  - a^2(t) (dx^2+dy^2+dz^2) \, .
$$
For $a \approx   t^p$, especially, if 
$p \ge 1$, this is called power-law inflation. 
For the Lagrangian $L = R^m$ with
 $ p=   - (m - 1) (2m - 1)/(m - 2)$ power-law 
inflation is an exact solution, as it
 is for Einstein gravity with a 
minimally coupled scalar field $\Phi$ in an exponential potential 
 $V(\Phi) = \exp (\mu \Phi) $ and also for the higher-dimensional Einstein 
equation with a special Kaluza-Klein ansatz. The synchronized coordinates 
are not adapted to allow a closed-form solution, so we write
$$
ds^2=a^2 \left( Q^2 ( a) da^2 - dx^2 -dy^2-dz^2\right) \, .
$$
The general solutions reads $Q(a) =   (a^b + C)^{f/b}$ 
with free integration 
constant $C$ ($C = 0$ gives exact power-law inflation) and $m$-dependent 
values $b$
and $f$: $f = -2 +1/p$, $b =   (4m - 5)/(m   - 1)$. 
Finally, special solutions for the closed and open Friedmann model are found.

\medskip


\noindent 
AAA subject classification 161
\end{abstract}

\section{Introduction}

The de Sitter space-time
\be
ds^2=dt^2  - e^{2Ht} (dx^2+dy^2+dz^2) \, , \qquad H \ne 0
\ee
is the space-time being mainly discussed to represent 
the inflationary phase of cosmic evolution. 
Recently, a space-time defined by
\be
ds^2=dt^2  - \vert t \vert ^{2p} (dx^2+dy^2+dz^2) \, , \qquad p \ne 0
\ee
enjoys increasing interest for these discussions, too. 
Especially, eq. (2) with $p \ge 1$, $ t > 0$ is called 
power-law inflation; and with $p < 0$, $t < 0$ it is called polar inflation.

\bigskip

We summarize some differential-geometrical properties 
of both de Sitter and power-law/polar inflation 
in sct. 2 and show, from which kind of 
scale-invariant field equations they arise in sct. 3; 
we give the complete set of solutions for 
the spatially flat and special solutions for the closed and open Friedmann 
models in closed form for field equations following from the Lagrangian
\be
L = R^m 
\ee
in sct. 4, and discuss the results in the final sct. 5 under the 
point of view that power-law inflation is an attractor solution of the 
corresponding field equations.

\section{Differential--geometrical properties}

Eq. (2) defines a self-similar space--time: if we 
multiply the metric $ds^2$ by an arbitrary positive 
constant $a^2$, then the resulting  $d \hat s^2 = a^2  ds^2$ 
 is isometric to $ds^2$. 
Proof: We perform a coordinate transformation 
$\hat t = at$, $\hat  x = b(a, p) x$ \dots $\Box$ 
On the other hand, the de Sitter space-time eq. (1) is not 
self-similar. 
Proof: It has a constant non-vanishing curvature scalar.  $\Box$ 

\bigskip

Power-law inflation is intrinsically time-oriented. 
Proof: The gradient of the curvature scalar defines a temporal 
orientation. $\Box$ 
On the other hand, the expanding ($H > 0$) and the contracting ($H < 0$) 
de Sitter space-time can be transformed into each other by a 
coordinate transformation. Proof: Both of them can be transformed 
to the closed Friedmann universe with scale factor $\cosh(Ht)$,
 which is an 
even function of $t$. $\Box$
This is connected with the fact 
that eq. (2) gives a global description, whereas eq. (1) gives only a  
proper subset of the full de Sitter space--time.

\bigskip

For $p \to \infty$, eq. (2) tends to  eq. (1). Such a statement has 
to be taken with care, even for the case with real functions. Even more 
carefully one has to deal with space-times. The most often used 
limit ---  the Geroch-limit of space-times --- has the property that a 
symmetry (here: self-similarity) of all the elements of the sequence must 
also be a symmetry of the limit.

\bigskip

From this it follows that the Geroch limit 
of space-times (2) with $p \to \infty$  cannot be unique, 
moreover, it is just the one-parameter set (1) parametrized by 
arbitrary values $H > 0$.

\section{Scale--invariant field equations}

A gravitational field equation is called scale-invariant, if 
 to each solution $ds^2$ and to each positive constant $c^2$  
the resulting homothetically equivalent metric $d \hat s^2 = c^2  ds^2$
is also a solution. 
One example of such field equations is that one following 
from eq. (3). 
Moreover, no Lagrangian $L = L(R)$ gives
 rise to a scale-invariant field 
equation which is not yet covered by (3) already.

\bigskip

Secondly, for
\be
     L = R/16 \pi G -    \frac{1}{2} g^{ij} \Phi_{, i}  \Phi_{, j} 
 + V_0 \exp(\mu \Phi)
\ee
(let be $8 \pi G = 1$, henceforth) the homothetic transformation has 
to be accompanied
 by a suitable translation of $\Phi$ to ensure scale-invariance. 
For $\mu \ne  0$, the value of $V_0$ can be normed to 1, 0 or $- 1$.

\bigskip

A third example is the following: for the Kaluza-Klein ansatz
\be
 dS^2     = ds^2(x^i) + W(x^i) d\tau^2(x^\alpha)  
\ee
$(i,j = 0, \dots 3$; $\a, \b = 4, \dots N- 1$;
 $A, B = 0 \dots N- 1)$  
the $N$-dimensional Einstein equation $R_{AB} = 0$
 is scale-invariant. Here, we 
restrict to the warped product of the $(N - 4)$-dimensional internal 
space $d\tau^2$ with  
 4--dimensional space--time $ds^2$.

\bigskip

Let us consider the limits $m \to \infty$  and 
$m \to 0$ of eq. (3). One gets 
$L = \exp (R/R_0)$  and $ L = \ln (R/R_0)$, resp. 
Both of them 
give rise to a field equation which is 
not scale-invariant: a homothetic transformation 
changes also the reference value $R_0$.
 For the second case 
this means a change of the 
cosmological constant. So we have a similar result as before: the limits 
exist, but they are not unique.

\section{Cosmological  Friedmann models}

We consider the closed  and open model in sct. 4.1. 
and the spatially flat model in sct. 4.2.

\subsection{The closed and open models}

Let us start with the closed model. We restrict 
ourselves to a region where one has 
expansion, so we may use the cosmic 
scale factor as time-like coordinate:
\be
ds^2=a^2 \left( Q^2 ( a) da^2 - d\sigma^2 \right)\, \qquad Q(a) > 0\, ,
\ee
where
\be
d\sigma^2  = 
dr^2 + \sin^2 r d \Omega^2 \, , \qquad
    d\Omega^2 = d\theta^2 + \sin^2 \theta  d\psi^2 
\ee
is the positively curved 3-space of constant curvature. 
It holds: if the 00-component of 
the field equation (here: eq. (10) below) is fulfilled, then all other 
components 
are fulfilled, too. Such a statement holds true for all 
Friedmann models and ``almost all sensible field 
theories". With ansatz (6, 7) we get via
\be
R^0_0  = 3(a^{-4} Q^{-2} + a^{-3} Q^{-3} dQ/da)  
\ee
and
\be
R = 6(a^{-3}Q^{-3}dQ/da - a^{-2}) 
\ee
the result: the field equation following from the 
Lagrangian (3) is fulfilled for metric (6, 7), if and only if
\be
     mR  R^0_0 -R^2/2
   + 3m(m -  1) Q^{-2} a^{-3} dR/da = 0  
\ee
holds. For $m = 1$ (Einstein's theory), no solution exists. For all other 
values $m$, eqs. (8-10) lead to a second order equation for $Q(a)$.

\bigskip

We look now for solutions with vanishing $R^0_0$, 
 i.e., with (8) we get
\be
     Q= C/a \,  , \qquad   C= {\rm  const.} >0 \, . 
\ee
By the way, (6,7) is self-similar if and only 
if  eq. (11) holds. From (9, 11) we get
\be
     R=D/a^2\,  , \qquad     D=-6(l +1/C^2) \, .   
\ee
We insert (11, 12) into (10) and get
\be
C =  (2m^2 - 2m - 1)^{1/2}   \, ,
\ee
which fulfils (11) if
\be
     m >(1 + \sqrt 3)/2 \quad {\rm  or} \quad      
m < (1 - \sqrt 3)/2
\ee
holds. Inserting (11,13) into (6) and introducing synchronized 
coordinates we get as a result: if (14) holds, then 
\be
ds^2 = dt^2  - t^2 d\sigma^2 /  (2m^2 - 2m - 1)
\ee
is a solution of the fourth order field equation 
following from Lagrangian (3). It is a 
self-similar solution, and no other self--similar solution 
describing a closed Friedmann model exists.

\bigskip

For the open Friedmann model all things are 
analogous, one gets for
$$
 (1 - \sqrt 3)/2 < m < (1 + \sqrt 3)/2
$$
and with sinh $r$ instead of sin $r$ in eq. (7) the only self-similar open 
solution (which is flat for $m \in \{0, 1\}$)
$$
ds^2 = dt^2  - t^2 d\sigma^2 /  (- 2m^2 + 2m + 1) \, .
$$

\subsection{The spatially flat model}

The field equation for the spatially flat model 
can be deduced from that one of a closed model 
by a limiting procedure as follows: we 
insert the transformation $r \to \e r$, 
$  a \to   a/\e$, $  Q \to  Q\e^2$  
 and 
apply the limit $\e \to  0$ afterwards. One gets via
$$
\lim_{\e \to 0} \  \sin (\e r) = r  
$$
the metric
\be
ds^2=a^2 \left( Q^2 ( a) da^2 - dx^2 -dy^2-dz^2\right)
\ee
with unchanged eqs. (8, 10), whereas eq. (9) yields
\be
     R = 6a^{-3}Q^{-3} dQ/da \, .  
\ee
The trivial solutions are the flat Minkowski space-time 
and the model with constant value 
of $Q$, i.e., $R = 0$ ($m > 1$ only), which is simply Friedmann's 
radiation model.

\bigskip

Now, we consider only regions with non-vanishing curvature 
scalar. For the next step we apply the fact that the spatially flat
 model has one symmetry more than the closed one: 
the spatial part of the metric is self-similar. In the coordinates 
(16) this means that each 
solution $Q(a)$  may be multiplied by an arbitrary constant. 
To cancel this arbitrariness, we define a new function
\be
P(a) =  d(\ln  Q)/da \,  .
\ee
We insert (8, 17) into (10) and then (18) into the resulting 
second order equation for $Q$. We get the first order equation for $P$
\be
0    = m(m - 1) dP/da + (m - 1) (1 - 2m) P^2 + m(4 - 3m) P/a \, .
\ee
As it must be the case, for $m = 0,1$,  only $P = 0$  
is a solution. For $m = 1/2$, $P \sim  a^5$  and 
therefore, $Q = \exp(ca^6)$, $c$ denotes an 
integration constant. For the other values $m$ we define $ z =  aP$
 as new 
dependent and $t=\ln a$   as new independent variable.

Eq. (19) then becomes $0=  dz/dt + gz^2 - bz$,
 $g = l/m - 2$, $b = (4m -5)/(m-1)$. 
For $m = 5/4$ we get $ z = - 5/(6t - c)$
 i.e., $P = - 5/(6a \ln (a/c))$, $
Q=( \ln (a/c) )^{-5/6} $ \, .
 For the other values
 we get
\bea
    z =f/ (\exp(-bt + c) - 1) \, , \quad
f= -b/g\,  , \quad 
P =f/(e^c a^{1-b} - a)
\nonumber \\
Q = (\pm a^b + c)^{1/g} \, .
\eea
For $m \to
   1/2$  we get $b \to  6$  and 
$1/g \to \infty$; for $m \to 5/4$
 we get $b \to  0 $  and $1/g \to    -5/6$, so  the
 two special cases could also have 
been obtained by a limiting procedure from eq. (20).

\bigskip

Metric (16) with (20) can be explicitly written in synchronized 
coordinates for special examples only, see e.g. BURD and BARROW (1988).

\section{Discussion}

We have considered 
scale--invariant field equations. The three 
examples mentioned in sct. 3 can 
be transformed into each other by a 
conformal transformation of the 
four--dimensional space-time metric. 
The parameters of eqs. (3) and (4) are related 
by $ \sqrt 3 \,  \mu = \sqrt 2 (2 -m)/(m      - 1)$, 
cf. SCHMIDT (1989), 
a similar relation exists to the internal space dimension in
eq. (5), one has 
$$
m = 1 + 1/ \left\{
 1  + \sqrt 3
\left[
1 + 2/(N - 4)
\right]^{\pm 1/2}
\right\}  \, , 
$$
for 
details cf. BLEYER and SCHMIDT (1990). The 
necessary conformal factor is a suitable power 
of the curvature scalar (3). A further conformal transformation in 
addition with the field redefinition $\theta  = \tanh \Phi $ 
 leads to the 
conformally coupled scalar field $\theta$ 
in the potential
$$
 (1 +  \theta )^{2 +  \mu } \, (1 -  \theta )^{2 -  \mu } \, , 
$$
the conformal factor being $\cosh^4 \Phi$, 
cf. SCHMIDT (1988). So, equations 
stemming from quite different physical foundations 
are seen to be equivalent. 
We have looked at 
them from the point of view of  self-simi1ar 
solutions and of limiting processes changing the type of symmetry.

\bigskip

The general solution for the spatially 
flat Friedmann models in fourth order 
gravity (3), eqs. (16, 20), can be written for small $c$ in synchronized 
coordinates as follows
$$
ds^2=dt^2  - a^2(t) (dx^2+dy^2+dz^2) \, , \qquad
 a(t) = t^p ( 1 + \e t^{-bp} + O(\e^2)) \, . 
$$
$\e = 0$ gives exact power-law inflation 
with $p = (m - 1) (2m - 1)/(2 - m)$ and $b = (4m - 5)/(m - 1)$.
 We see: in the 
range $1 < m < 2$, power-law
 inflation is an attractor solution within the set 
of spatially flat Friedmann models for $m \ge  5/4$ only. The
generalization to 
polar inflation ($m > 2$, $p <0$) is similar.

\bigskip
\noindent 
{\it Acknowledgement.}
  Interesting discussions of  the
 topic with L. AMENDOLA, V. M\"ULLER, 
P. TEYSSANDIER, and A. A. STAROBINSKY are gratefully acknowledged.

\section*{References}

\noindent
BLEYER, U., SCHMIDT, H.-J.: 1990, Int. J. Mod. Phys. A, in print.

\noindent
BURD, A., BARROW, J.: 1988, Nuc1. Phys. B {\bf  308}, 929.

\noindent
SCHMIDT, H.-J.: 1988, Phys. Lett. B {\bf 214}, 519.

\noindent
SCHMIDT, H.-J.: 1989, Class. Quant. Grav. {\bf 6}, 557.

\bigskip

\noindent 
Received 1989 November 7

\bigskip

\noindent 
 {\small { In this reprint done with the kind permission of the 
copyright owner 
we removed only obvious misprints of the original, which
was published in Astronomi\-sche Nachrichten:   
 Astron. Nachr. {\bf 311} (1990) Nr. 3, pages 165 - 168.
 The first cited reference appeared in: 
  Int. J. Mod. Phys. A {\bf  5}
(1990)  4671 - 4676.

\bigskip

\noindent 
  Author's address  that time: 

\noindent
H.-J. Schmidt,  
Zentralinstitut f\"ur  Astrophysik der AdW der DDR, 
1591 Potsdam, R.-Luxemburg-Str. 17a
}}
\end{document}